\documentclass[letterpaper, 10 pt, conference]{ieeeconf}  

\IEEEoverridecommandlockouts                              
\usepackage{graphics} 
\usepackage{multirow}
\usepackage[table,xcdraw]{xcolor}
\usepackage{graphicx}
\usepackage{float}

\makeatletter
\def\normaljustify{%
  \let\\\@centercr\rightskip\z@skip \leftskip\z@skip%
  \parfillskip=0pt plus 1fil}
\makeatother

\title{\LARGE \bf
Multi-modal Approach for Affective Computing
}

\author{Siddharth$^{1,2}$, Tzyy-Ping Jung$^{2}$ and Terrence J. Sejnowski$^{2}$
\thanks{$^{1}$Siddharth is with the Electrical and Computer Engineering Department at University of California San Diego, California, USA
        {\tt\small ssiddhar@eng.ucsd.edu}}%
\thanks{$^{2}$Tzyy-Ping Jung and Terrence J. Sejnowski are with the Institute for Neural Computation, University of California San Diego, California, USA
        {\tt\small jung@sccn.ucsd.edu, terry@salk.edu}}%
}

\begin{document}

\maketitle
\thispagestyle{empty}
\pagestyle{empty}

\begin{abstract}
Throughout the past decade, many studies have classified human emotions using only a single sensing modality such as face video, electroencephalogram (EEG), electrocardiogram (ECG), galvanic skin response (GSR), etc. The results of these studies are constrained by the limitations of these modalities such as the absence of physiological biomarkers in the face-video analysis, poor spatial resolution in EEG, poor temporal resolution of the GSR etc. Scant research has been conducted to compare the merits of these modalities and understand how to best use them individually and jointly. Using multi-modal AMIGOS dataset, this study compares the performance of human emotion classification using multiple computational approaches applied to face videos and various bio-sensing modalities. Using a novel method for compensating physiological baseline we show an increase in the classification accuracy of various approaches that we use. Finally, we present a multi-modal emotion-classification approach in the domain of affective computing research.
\end{abstract}

\section{INTRODUCTION}
Affective Computing is a well-established research field covering a wide spectrum of applications from monitoring health to human-computer interaction. Majority of studies on affective computing have used vision-based approaches \cite{face_only,face_AUs}, where a camera monitors subject's facial expressions and classifies them to one of the pre-defined emotion categories. These vision-based approaches lack a basic understanding of the physiological state of the subject and only take into account the facial expression at that moment which may be independent of the actual emotional state. On the other hand, some studies that do employ bio-signal based methods suffer two limitations: not defining a proper physiological baseline for the subject i.e. a measure of emotional state at the start of the experiment \cite{affective_computing_survey,GSR_and_PPG} and the use of a single bio-sensing modality with its inherent limitations \cite{face_only,affective_computing_survey}. Furthermore, a performance comparison between these various methods is largely unexplored. The work presented in this paper addresses these issues by presenting a novel method to assess emotional baselines for various sensing modalities. It then goes on to combine the modalities together for generating more robust models than those by individual modalities. In the path to carry out above objectives this work also presents a comparative study of various techniques and algorithms used for this multi-modal analysis.

Use of affective computing for assessing emotions when people are watching a content on a screen or in virtual reality is driven by many motivations. Content rating for advertisement purposes is an old area of research, but the development of wearable multi-modal bio-sensing hardware systems \cite{my_hcii} has added another dimension to it. Moreover, in the absence of noise arising due to motion in real-world situations, such experimental setups provide easy and quick ways to develop affective computing framework. Such studies do not take a holistic view of all the bio-sensing modalities along with those of vision. Such studies usually either only work in the multi-modal domain of video/audio/text \cite{multi-modal-without-biosensing} or in the multi-modal bio-sensing domain of EEG/GSR etc. \cite{GSR_and_PPG}. They do not combine the features from these two domains. The emotional state of a subject when s/he enters to begin an experiment is of very high importance since it directly influences how s/he reacts to the emotional stimuli. For example, no matter what content a subject sees, s/he will not be able to rate a content very positive if s/he has suffered from a negative event like the death of a pet animal not long before the experiment. This makes emotional baseline compensation of great importance as also highlighted by our results below. In this research domain, most of the results for various emotional measures have reported accuracies below 70\% \cite{affective_computing_survey}. We present here novel methods for feature extraction and baseline compensation to perform much better than the results reported previously.

To briefly outline the procedure: We used EEG, ECG, GSR, and frontal videos of subjects when they are watching emotionally stimulating videos being displayed on a screen. For each subject and trial (video), we have the valence and arousal ratings reported by the subject before and after every experimental trial. For each of the above modalities, we evaluate the performance with and without baseline compensation. We then predict their arousal, valence, liking, dominance, and emotional state (four/eight classes) using the rest of the database. To the best of our knowledge, this is the most holistic framework that uses such widely varying modalities with computation and compensation for the emotional baseline. Another important aspect of our work is that we show state-of-the-art results despite the dataset containing trials with varying time lengths with the shortest being only a third of the longest one.

\section{DATASET DESCIPTION AND PROPOSED METHOD}
\begin{figure}[H] \centering
{\includegraphics[width=2.8in, height = 2.2in]{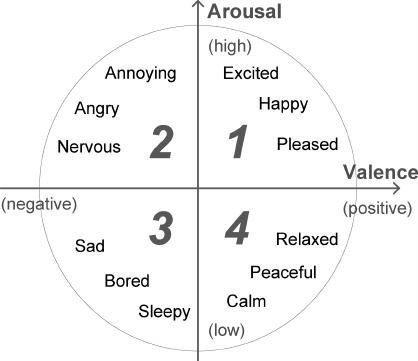}}

\normaljustify 
\caption{Emotion Circumplex Model}
\label{ref:fig1}
\end{figure}

This study uses AMIGOS Dataset \cite{amigos} for designing and evaluating our framework. This dataset has 40 subjects watching 16 short (51-150 seconds) video clips (total 640 video trials). The videos have been taken from other generic emotion-based datasets and have been annotated by external participants too. EEG (14 channels), ECG (2 channels), GSR (1 channel), and frontal video (RGB) of the subjects were recorded simultaneously during the experiments. EEG was sampled at 128 Hz, ECG and GSR have been down-sampled from 256 Hz to 128 Hz, whereas the videos were captured at 25 frames per second (fps). For each trial, the subject's self-reported valence, arousal, liking, and dominance before and after the trial are also recorded on a scale of 1 to 9. We use Russell's circumplex model \cite{russell} to assess emotional state for these valence and arousal values (Fig. \ref{ref:fig1}). Valence corresponds to the ``mood" of the subject while watching the videos i.e. positive or negative mood, whereas Arousal corresponds to the ``arousing" state ranging from low to high. Similarly, Liking and Dominance measures assess how much the subject likes the content and how much s/he dominates the emotional stimuli being witnessed respectively. For clarity, we divide each of the extraction of features from all the modalities described above into subsections below. Each subsection consists of the methods we use for extracting features for that modality. Finally, we present the method for baseline correction used by us.

\subsection{EEG-based feature extraction}
We employ two distinct methods for extracting features from the EEG data. 

\subsubsection{Conditional entropy features} 
For each of the possible pairs of the 14 EEG sensors, we compute the conditional entropy of the EEG signal for every trial. This measure contains information about the amount of information present in one signal given the other signal. In this manner, we have 91 EEG features for all possible channel pairs using \cite{conditional_entropy}.

\subsubsection{CNN-based features using the EEG topography}
\begin{figure}[H] \centering
{ \includegraphics[width=3.2in, height = 0.8in]{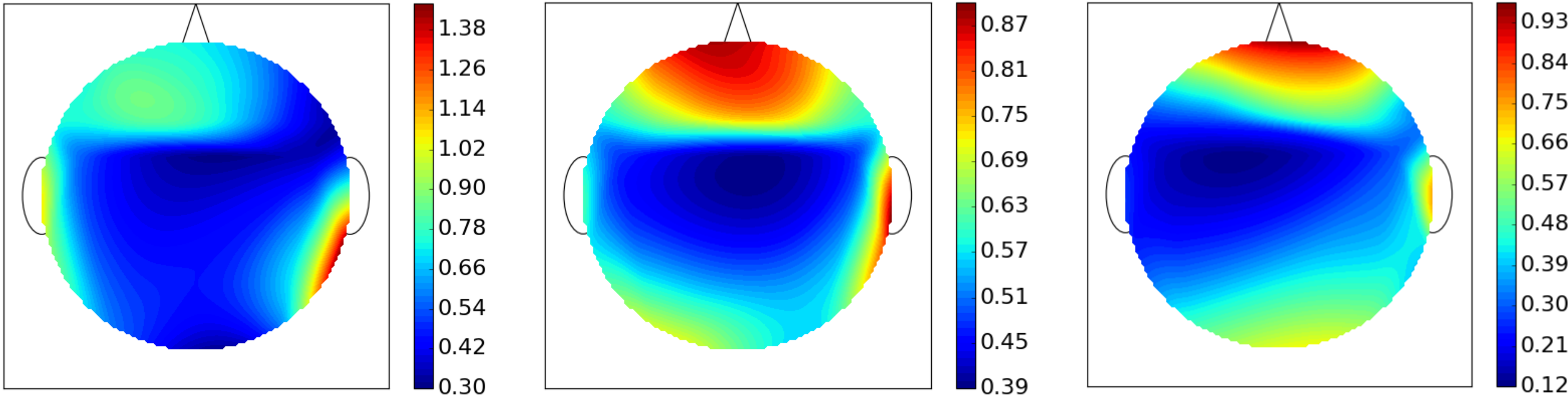}}

\normaljustify 
\caption{Theta, Alpha and Beta band baseline PSD ratio with topography information (color bar, boundary box and face parts have been added for visualization only)}
\label{ref:fig2}
\end{figure}

We compute power spectral density (PSD) with half-second overlapping windows for every EEG channel in the theta (4-7 Hz), alpha (7-13 Hz), and beta (13-30 Hz) bands over the EEG data. The PSD for every second is then averaged over the length of the video.  The raw PSD values so obtained do not work very well in the emotion-classification problem \cite{amigos}. Hence, we tried to use convolutional neural networks (CNNs) for extracting more useful features based on the PSD values. To use the topography information of the EEG, we project the PSD features onto a 2-D brain heatmap (Fig. \ref{ref:fig2}) using the EEG 10-20 system. We then used the pre-trained VGG-16 network \cite{VGG}, which has been trained on more than a million images. It has been shown that off-the-shelf features extracted from this deep neural network can also perform well for various tasks \cite{VGG_2}. The PSD images were then resized to $224{\times}224$ pixels, which is the size of input images for the VGG-16 network. We then extract 4,096 deep CNN-VGG features, which are the output of the last max pooling layer of the CNN network from these PSD heatmaps for each of the three EEG bands. We then used Principal Component Analysis (PCA) to reduce the number of features to the 32 most representative ones for each band, resulting in a total of 96 features. We then concatenated the EEG features from the above two techniques.

\subsection{ECG-based feature extraction}
\begin{figure}[H] \centering
{\includegraphics[width=3.2in, height = 1.2in]{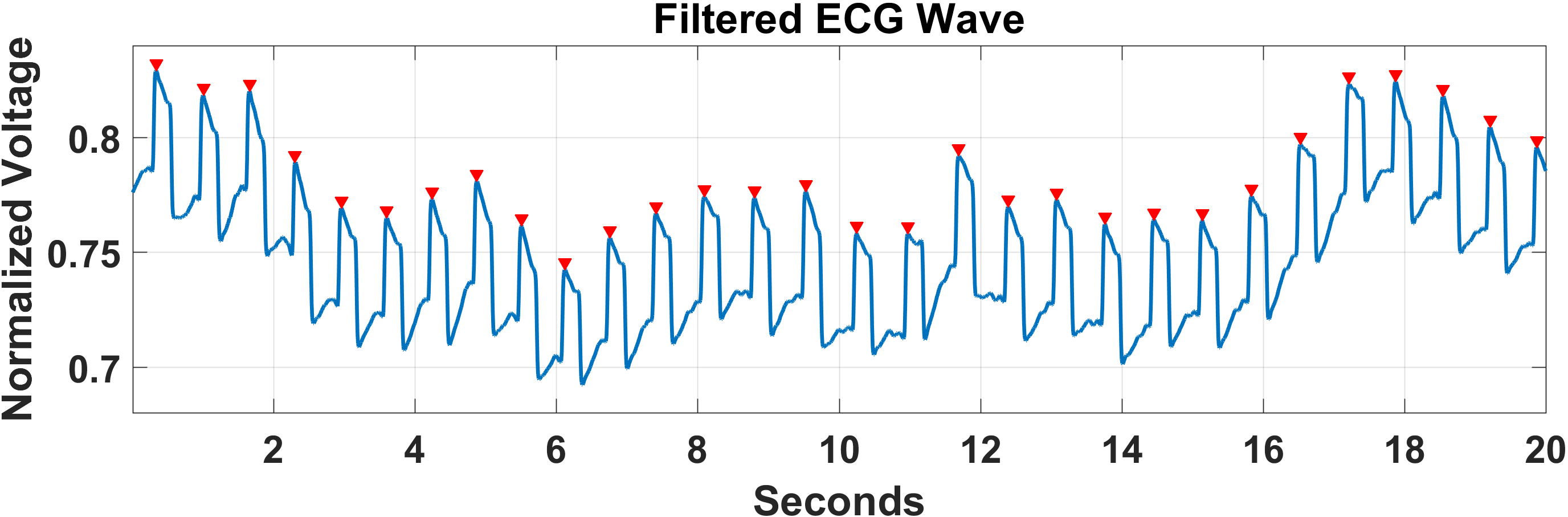}}

\normaljustify 
\caption{Inter-beat (RR) intervals were computed as distance between successive peaks (marked in \textcolor{red}{red}) in the ECG wave to calculate HRV.}
\label{ref:fig3}
\end{figure}

We first cleaned the noise peaks in the ECG data using a moving-average filter with a window length of 0.25 second after which we used a peak-detection algorithm to find the peaks in ECG. A threshold on the minimum time difference between successive peaks to be at least 0.5 seconds was imposed to remove false ECG T-wave peaks. Heart Rate Variability (HRV) has been shown as a reliable measure for assessing emotional states \cite{hrv_only}. We exploited two ECG channels for computing HRV using the pNN-50 algorithm \cite{pNN50_algorithm} based on the duration of inter-beat (RR) intervals and use them as features (Fig. \ref{ref:fig3}).

\subsection{GSR-based feature extraction}
Similar to the ECG analysis, we employed a moving-average filter to clean the data. GSR is a single-channel signal measuring the changes in the voltage potentials on the skin. Not being very descriptive in the high-frequency range like EEG or fast changing like ECG, we compute eight time-domain statistical features from the GSR signals. 

\begin{table*}[!ht]
\centering
\caption{Individual Modality Performance}
\label{table-single-modality}
{%
\begin{tabular}{c|c|c|c|c|c|c|c|c|c|c|}
\cline{2-11}
\multirow{2}{*}{}                                  & \multicolumn{5}{c|}{\textbf{Without Baseline Compensation}}                    & \multicolumn{5}{c|}{\textbf{Baseline Compensated}}                             \\ \cline{2-11} 
                                                   & \textbf{EEG} & \textbf{ECG} & \textbf{GSR} & \textbf{F.V.-1} & \textbf{F.V.-2} & \textbf{EEG} & \textbf{ECG} & \textbf{GSR} & \textbf{F.V.-1} & \textbf{F.V.-2} \\ \hline
\multicolumn{1}{|c|}{\textbf{Valence}}             & 65.04        & 58.73        & 61.72        & 67.97           & 60.94           & 66.67        & 58.73        & 64.84        & 68.75           & 65.63           \\ \hline
\multicolumn{1}{|c|}{\textbf{Arousal}}             & 69.92        & 57.94        & 57.81        & 72.66           & 72.66           & 71.54        & 57.94        & 63.28        & 72.66           & 76.56           \\ \hline
\multicolumn{1}{|c|}{\textbf{Liking}}              & 64.23        & 69.05        & 64.06        & 71.09           & 66.41           & 70.69        & 69.49        & 70.00        & 69.17           & 73.33           \\ \hline
\multicolumn{1}{|c|}{\textbf{Dominance}}           & 69.92        & 55.56        & 59.38        & 70.31           & 72.66           & 72.36        & 57.14        & 57.81        & 73.44           & 71.88           \\ \hline
\multicolumn{1}{|c|}{\textbf{Emotion (4 Classes)}} & 44.72        & 28.57        & 33.59        & 49.22           & 46.09           & 48.78        & 34.13        & 32.81        & 46.09           & 48.44           \\ \hline
\multicolumn{1}{|c|}{\textbf{Emotion (8 Classes)}} & 34.96        & 28.57        & 25.00        & 30.47           & 30.47           & 34.15        & 24.60        & 24.22        & 28.91           & 38.28           \\ \hline
\end{tabular}
}
\vspace{1ex}

\normaljustify         
F.V.-1: Face localization features, F.V.-2: CNN-VGG features. Baseline compensation almost always allows for better performance.
\end{table*}

The first two features are the number of peaks and the absolute mean height of the peaks in the signal. Six statistical features are then computed as described in \cite{gsr_features}.

\subsection{Video-based feature extraction}
We extract a single frame for every second from each of the frontal videos to save computation. As in the EEG analysis, we use the frontal videos to compute two distinct types of features.

\subsubsection{Face localization features}
\begin{figure}[H] \centering
{\includegraphics[width=1.8in, height = 1.5in]{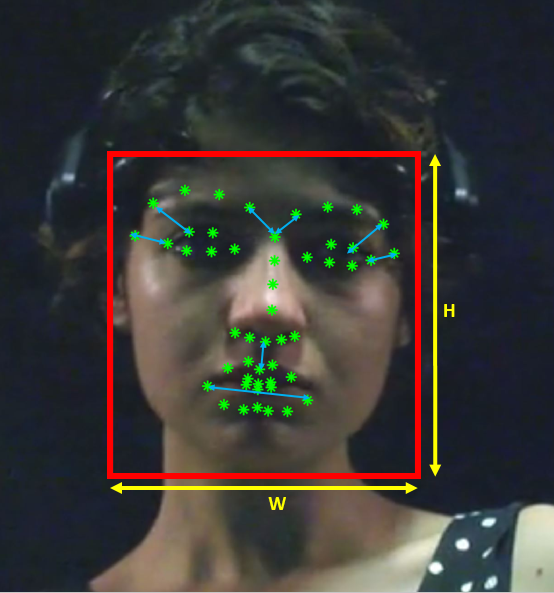}}

\normaljustify
\caption{Face localized points marked in \textcolor{green}{green} and a subset of features used to calculate face AUs marked as \textcolor{cyan}{cyan} lines. Height (H) and width (W) of the face were used to normalize the features. Subject's consent to use her face for publication is marked in the dataset.}
\label{ref:fig4}
\end{figure}

We use state-of-the-art Chehra \cite{chehra} algorithm to compute 49 face localized points from the image frame (Fig. \ref{ref:fig4}). We then use these localized face points to extract 30 features based on facial Action Unit (AU) recognition \cite{face_AUs}. These features include some self-designed features such as those that take into account the distance between the eyebrow from the eye, between the centers of upper and lower lip, etc. These features are then scaled as per the dimensions of the face in the image to account for the varying distance from the camera. We believe that such features for AU classification are highly representative of the human emotion as presented in other studies \cite{multi-modal-without-biosensing}. For all selected video frames in a single video, we calculate the mean, $95^{th}$ percentile, and standard deviation of these features so that these combined 90 features (30 each for mean, $95^{th}$ percentile, and standard deviation) represent the complete video and not just an individual video frame.

\subsubsection{CNN based features}
We use the deep CNN-VGG network like we did for the EEG topography images above on all video frames to calculate 4096 features with their mean, $95^{th}$ percentile, and standard deviation across the frames in a trial. We then down-sampled this feature space to 50 using PCA. The only difference being we utilized a version of the network trained only on face images \cite{VGG_Faces} to extract features more relevant to faces.

\subsection{Baseline compensation method}
For each affective response (valence, arousal, liking, and dominance), the rating given by the subjects on the scale of 1 to 9 before and after each trial were used for compensating the baseline for that trial. This was done by first computing the difference in the value before the trial and expected (neutral) baseline value i.e. 5. The difference thus calculated was added to the response value after the trial. Hence, for example, if a subject reports valence of 9 i.e. being in extremely positive ``mood" before a trial, the deviation from neutral is 4. Now, if this subject reports his/her valence after the trial as 2, the baseline compensated valence would be 6. This is because the subject would be physiologically incapable of actually feeling an emotional response at the opposite end of the spectrum in that short time interval.

\section{EVALUATION}
We randomly divide the dataset to training and testing sets with an 80/20\% distribution i.e. 512 trials for training and 128 for testing. This was done to generate a more robust model for cross-subject performance rather than taking out an equal number of samples from all the subjects. If for any trial the data for a modality was missing in the dataset, then it was omitted. For each of the above modalities, we use extreme learning machine (ELM) \cite{ELM} for classification with 10-fold cross-validation and a sigmoid activation function for training.

We divide each of the valence, arousal, liking, and dominance values into two (high/low) classes. Similarly, we divide the emotion classes into four and eight quadrants using the circumplex model defined above to evaluate our algorithm. The eight quadrants loosely refer to pleased, excited, annoying, nervous, sad, sleepy, calm, and relaxed emotions. These classes were formed for both without and with baseline compensation.

\subsection{Individual Modality and Baseline Compensation Evaluation}
In this section, we evaluate the methods we developed above for various modalities. We also evaluate the effect of the baseline compensation on the classification performance for each of the above mentioned modalities. Table \ref{table-single-modality} shows the results of classification accuracy on various classes for all the modalities with and without compensating for the baseline. It is clear from Table \ref{table-single-modality} that in almost all cases and for all modalities the accuracy increases after baseline compensation. In general, EEG performs best among the three bio-sensing modalities and at times perform as well as video-based analysis. Furthermore, the high accuracies shown by all methods for liking and dominance indicate that the responses recorded for these by participants are highly correlated to their physiology i.e. even more than for valence and arousal. This is intuitive since it is easier to rate liking and such parameters for a video than to rate valence and arousal which are subjective depending highly on the emotional baseline. It is evident from a recent affective computing survey \cite{affective_computing_survey} that the results presented here beat the accuracies for all kinds of responses including that for four- and eight-class emotion classification. For all the cases, the accuracy obtained by our proposed methods are sufficiently greater than the chance accuracy.

The CNN-VGG based features for both EEG and frontal videos show an increase in classification performance after compensating baseline, suggesting that they are able to extract features representing physiological variations very well. It is also interesting to note that for eight-class emotion classification problem, the accuracy actually decreases for bio-sensing based methods. We believe this is due to the scale between 1 to 9 for valence and arousal having a small range for classification into eight emotion classes.

\begin{table}[!ht]
\centering
\caption{Multi-modality Fusion Performance}
\label{table-multi-modality}
{%
\begin{tabular}{|c|c|}
\hline
\textbf{Method}                    & \textbf{Emotion Classification Accuracy} \\ \hline
EEG + Frontal videos & 52.51\%                                         \\ \hline
GSR + ECG      & 38.28\%                                         \\ \hline
\end{tabular}
}
\vspace{1ex}

\normaljustify         
Using multiple modalities allow for better emotion classification performance for four emotion classes.
\end{table}

\subsection{Combining Modalities Evaluation}
As pointed out above in Introduction, not many studies focus on more than one modality in affective computing research. Hence, another goal of this study was to increase the emotion classification accuracy while fusing features from different modalities. Table \ref{table-multi-modality} shows the results for two such cases. We used the baseline compensated emotion response values for this evaluation.

In the first case, we combined the features from EEG- and frontal-video-based methods and were able to increase the accuracy to 52.51\% as compared to these two methods individually. Similarly, we combined the features from GSR and ECG to increase the performance of these bio-sensing modalities. We get 38.28\% accuracy by the fusion of features from GSR and ECG which is greater than the accuracy we achieved by these modalities individually. Combining different modalities in this manner, we obtained similar results for other emotional responses such as valence, arousal, and liking.

\section{CONCLUSION}
This study proposed a new approach for the computation of physiological baseline while classifying human emotions, valence, arousal, liking, and dominance for camera-based and various bio-sensing modalities. We showed that baseline compensation improves classification performance and further we can use such modalities either independently or jointly for emotion classification. In the process, we also compared the performance of various modalities using feature-extraction methods not evaluated previously. We intend to take our work forward by acquiring more data with a different experimental setup such as that involving motion and ``real-world" emotional stimuli to make our algorithms more robust and scalable.

\addtolength{\textheight}{-12cm}   







\begin{thebibliography}{99}

\bibitem{face_only} Bailenson, J.N., Pontikakis, E.D., Mauss, I.B., Gross, J.J., Jabon, M.E., Hutcherson, C.A., Nass, C. and John, O., 2008. Real-time classification of evoked emotions using facial feature tracking and physiological responses. International journal of human-computer studies, 66(5), pp.303-317.

\bibitem{face_AUs} Tian, Y.I., Kanade, T. and Cohn, J.F., 2001. Recognizing action units for facial expression analysis. IEEE Transactions on pattern analysis and machine intelligence, 23(2), pp.97-115.

\bibitem{affective_computing_survey} Alarcao, S.M. and Fonseca, M.J., 2017. Emotions recognition using EEG signals: a survey. IEEE Transactions on Affective Computing.

\bibitem{GSR_and_PPG} Udovičić, G., Đerek, J., Russo, M. and Sikora, M., 2017, October. Wearable Emotion Recognition System based on GSR and PPG Signals. In MMHealth 2017: Workshop on Multimedia for Personal Health and Health Care.

\bibitem{my_hcii} Patel, A., Jung, T.P. and Sejnowski, T.J., 2017, July. An Affordable Bio-Sensing and Activity Tagging Platform for HCI Research. In International Conference on Augmented Cognition (pp. 399-409). Springer, Cham.

\bibitem{multi-modal-without-biosensing} Poria, S., Cambria, E., Hussain, A. and Huang, G.B., 2015. Towards an intelligent framework for multimodal affective data analysis. Neural Networks, 63, pp.104-116.

\bibitem{amigos} Miranda-Correa, J.A., Abadi, M.K., Sebe, N. and Patras, I., 2017. AMIGOS: A dataset for Mood, personality and affect research on Individuals and GrOupS. arXiv preprint arXiv:1702.02510.

\bibitem{russell} Russell, J.A., 1978. Evidence of convergent validity on the dimensions of affect. Journal of personality and social psychology, 36(10), p.1152.

\bibitem{conditional_entropy} Peng, H., Long, F. and Ding, C., 2005. Feature selection based on mutual information criteria of max-dependency, max-relevance, and min-redundancy. IEEE Transactions on pattern analysis and machine intelligence, 27(8), pp.1226-1238.

\bibitem{VGG} Simonyan, K. and Zisserman, A., 2014. Very deep convolutional networks for large-scale image recognition. arXiv preprint arXiv:1409.1556.

\bibitem{VGG_2} A. S. Razavian, H. Azizpour, J. Sullivan, and S. Carlsson, “CNN Features off-the-shelf: an Astounding Baseline for Recognition”, arXiv:1403.6382v3, 2014.

\bibitem{hrv_only} Orini, M., Bailón, R., Enk, R., Koelsch, S., Mainardi, L. and Laguna, P., 2010. A method for continuously assessing the autonomic response to music-induced emotions through HRV analysis. Medical \& biological engineering \& computing, 48(5), pp.423-433.

\bibitem{pNN50_algorithm} Ewing, D.J., Neilson, J.M. and Travis, P.A.U.L., 1984. New method for assessing cardiac parasympathetic activity using 24 hour electrocardiograms. Heart, 52(4), pp.396-402.

\bibitem{gsr_features} Mera, K. and Ichimura, T., 2004. Emotion analyzing method using physiological state. In Knowledge-Based Intelligent Information and Engineering Systems (pp. 195-201). Springer Berlin/Heidelberg.

\bibitem{chehra} Asthana, A., Zafeiriou, S., Cheng, S. and Pantic, M., 2014. Incremental face alignment in the wild. In Proceedings of the IEEE Conference on Computer Vision and Pattern Recognition (pp. 1859-1866).

\bibitem{ELM} Huang, G.B., Zhu, Q.Y. and Siew, C.K., 2006. Extreme learning machine: theory and applications. Neurocomputing, 70(1-3), pp.489-501.

\bibitem{VGG_Faces} Parkhi, O.M., Vedaldi, A. and Zisserman, A., 2015, September. Deep Face Recognition. In BMVC (Vol. 1, No. 3, p. 6).

\end{thebibliography}
\end{document}